\documentclass{article}
\usepackage[T1]{fontenc}
\usepackage{a4wide}
\usepackage{color}
\usepackage{graphics}

\makeatletter

\providecommand{\LyX}{L\kern-.1667em\lower.25em\hbox{Y}\kern-.125emX\@}

\newcommand{\lyxrightaddress}[1]{
  \par {\raggedleft \begin{tabular}{l}\ignorespaces
  #1
  \end{tabular}
  \vspace{1.4em}
  \par}
}

\usepackage[T1]{fontenc}

\makeatletter

\usepackage[T1]{fontenc}

\makeatletter

\usepackage[T1]{fontenc}

\makeatletter

\usepackage[T1]{fontenc}

\makeatletter

\usepackage[T1]{fontenc}


\makeatother
\makeatother
\makeatother
\makeatother

\begin{document}

\title{Charged Particle Fluctuation in Heavy Ion Collisions\thanks{
Talk given at the CRACOW EPIPHANY CONFERENCE ON QUARKS AND GLUONS IN EXTREME
CONDITIONS, 3 - 6 January 2002, Cracow, Poland
}}

\author{\underbar{Fritz W. Bopp} and Johannes Ranft \\
   Universität Siegen, Fachbereich Physik, \\
D--57068 Siegen, Germany \\
}

\maketitle
\vspace*{-8cm}

\lyxrightaddress{SI-02-2}

\vspace*{+7cm}

\begin{abstract}
Comparing quantities to analyze charged fluctuations in heavy ion experiments
the dispersion of the charges in a central rapidity box was found to be best
suited. Various energies and different nuclear sizes are considered in an explicit
Dual-Parton-Model calculation using the DPMJET code and a randomized modification
to simulated charge equilibrium. For large enough detection regions charged
particle fluctuations can provide a signal of the basic dynamics of heavy ion
processes. 
\end{abstract}

\section{{\large Charge Fluctuations in Fixed Target Hadron-Hadron Experiments }\large }

Let us look for a moment back to the analysis of purely hadronic multi-particle
production. At fixed target experiments it was possible to measure the charges
of all forward particles. In this way significant results could be obtained
with low energies available at the seventies~\cite{idschok73,bialas75,whitmore76,foa75,brandelik81}:

\begin{itemize}
\item {\small The charge fluctuations found involve mostly a restricted rapidity range. }{\small \par}
\item {\small Good agreement was obtained with cluster models. }{\small \par}
\end{itemize}
{\par\raggedright {\small The Quigg-Thomas relation}~{\small \cite{quigg73,quigg75}
(assuming neutral clusters) for fluctuation across a rapidity \( y \) boundary
\begin{equation}
\label{eq-1}
<\delta Q_{>y}^{2}>=<(Q_{>y}-<Q_{>y}>)^{2}>=c\cdot dN^{\mathrm{non}\, \mathrm{leading}}_{\mathrm{charge}}/dy
\end{equation}
 was found to be roughly satisfied}~{\small \cite{bopp78,misra83,ranft75,chouyang73,sivers74,phua77,lam77,chiu76,diasdedeus77,jezabek77}. }
\begin{figure}
{\par\centering \resizebox*{!}{0.2\textheight}{\includegraphics{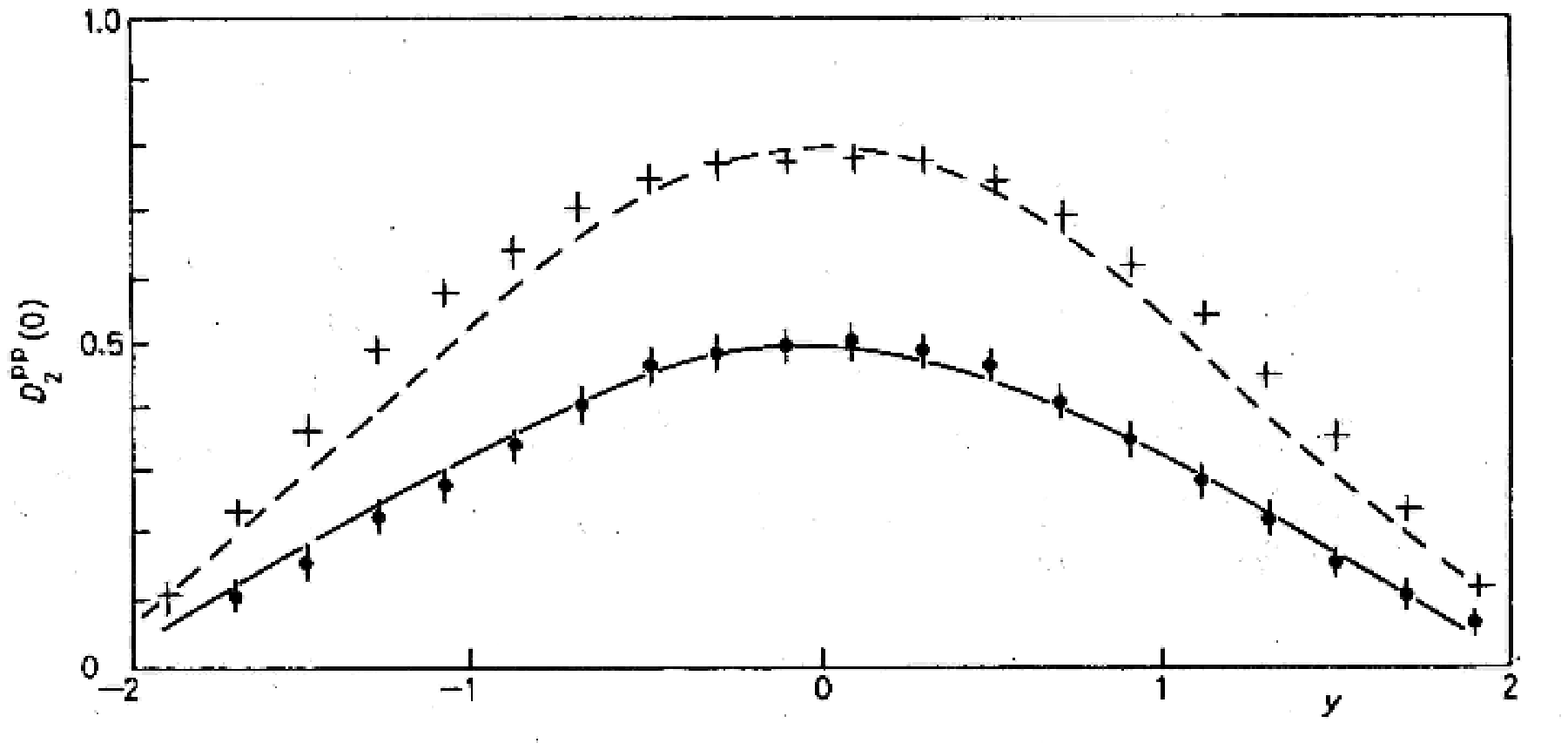}} \par}

\caption{The measured dispersion (+) and the produced charge dispersion ({*}), which
is corrected for leading charge flow, is compared with the suitably normalized
negative (produced) particle spectrum at 24 GeV/c.}
\end{figure}
{\small To illustrate, how early this relation was usefull we consider} \textsc{\textcolor{black}{\emph{\small 24
GeV proton proton scattering data.}}} {\small Taking the rapidity of the forward-backward
border as variable the data could be presented as in figure~1} ~{\small \cite{bialas75}.
The top and bottom lines and points correspond to with and without correcting
for the leading charge}\textbf{\emph{\small s}}{\small . The lines correspond
to suitably normalised spectrum. }\small \par}

{\par\raggedright {\small The agreement could be improved when string type \( q\bar{q} \)
-charge exchanges between the clusters were added}~{\small \cite{bopp78}.}
{\small Such exchanges appear in a large class of models. Using the Dual Parton
Model code DPMJET we re-checked the old results. For \( pp \)-scattering at
laboratory energies of 205 GeV good agreement was still obtained. The Quigg-Thomas
relation is satisfied with \( c=0.70 \) comparing to the experimentally preferred
value: \( c=0.72 \). }\small \par}

\section{{\large Charge Fluctuations in Heavy-Ion Scattering Experiments}\large }

In heavy ion scattering it is a central question whether the charges are distributed
just randomly or whether there is some of the initial dynamics left influencing
the global flow of quantum numbers. The charge flow measurements could again
be decisive. It is not an impractical conjecture. In heavy ion experiments the
charge distribution of the particle contained in a central box with a given
rapidity range \( [-y_{\mathrm{max}.},+y_{\mathrm{max}.}] \) as shown in figure~2 
\begin{figure}
{\par\centering \resizebox*{0.5\textwidth}{!}{\includegraphics{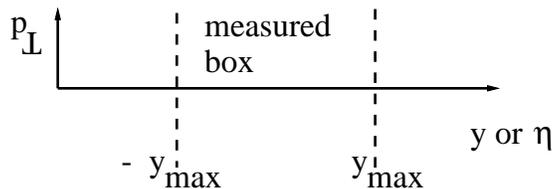}} \par}

\caption{Kinematic region of the ``central box''}
\end{figure}
can be measured and the dispersion of this distribution \( <\textrm{ }\delta Q^{2}> \)
can be obtained to sufficient accuracy. For sufficiently large gaps this quantity
contains information about long range charge flow. In comparison to the fluctuations
in the forward backward charge distributions the charge distribution into a
central box (having two borders) can be expected to require roughly twice the
rapidity range.

\section{{\large E}quilibrium {\large Expectations}\large }

Within the framework of equilibrium models it was proposed to use the quantity
to distinguish between particles emerging from an equilibrium quark-gluon gas
or from an equilibrium hadron gas~\cite{heinz00,jeon00,jeon99}. For a small
enough box in a central region at high energies where average charge flow can
be ignored, the (essentially) poisson distributed \emph{hadron gas} yields a
simple relation 

\begin{equation}
\label{eq-2}
<\delta Q^{2}>=<N_{\mathrm{charged}}>.
\end{equation}
 for any thermalised particles with charges \( 0 \) and \( \pm 1 \). The inclusion
of resonances reduces hadron gas prediction by a significant factor taken ~\cite{jeon00,bleicher00}
to be around \( 0.7 \). 

It is argued in the cited papers that this relation would change in a \emph{quark
gluon} gas to
\begin{equation}
\label{eq-3}
<\delta Q^{2}>=\sum _{i}q_{i}^{2}<N_{i}>=0.19<N_{\mathrm{charged}}>
\end{equation}
where \( q_{i} \) are the charges of the various quark species and where again
a central region is considered. The coefficient on the right was calculated~\cite{jeon00}
with suitable assumptions. A largely empirical final charged multiplicity \( N_{\mathrm{charged}}=\frac{2}{3}(N_{\mathrm{glue}}+1.2N_{\mathrm{quark}}+1.2N_{\mathrm{antiquark}}) \)
was used. 

It should be pointed out that the estimate is not without theoretical problems~\cite{bopp81,bialas02}.
There is also a number of systematic uncertainties in the above comparison.
As explained below in a simple approximation the result strongly depends on
what one takes as primordial particles and how the extra quarks needed for hadronization
are modelled. Considering these uncertainties we follow the conclusion of Fia{\l }kowski's
papers ~\cite{fialkowski00} that a clear cut distinction between the hadron-
and the quark gluon gas is rather unlikely. This does not eliminate the interest
in the dispersion as a \emph{measure of equilibration}.

\section{{\large Various Measures for Charge Fluctuations }\large }

\vspace{0.3cm}
For the analysis of the charge structure several quantities were discussed in
the recent literature. Besides the classic charge dispersion 
\begin{equation}
\label{eq-4}
<\delta Q^{2}>=<(Q-<Q>)^{2}>
\end{equation}
it was proposed to just measure the mean standard deviation of the ratio \( R \)
of positive to negative particles

\begin{equation}
\label{eq-5}
<\delta R^{2}>=<\left( \frac{N_{+}}{N_{-}}\: -<\frac{N_{+}}{N_{-}}>\right) ^{2}>
\end{equation}
or the quantity \( F \) 
\begin{equation}
\label{eq-6}
<\delta F^{2}>=<\left( \frac{Q}{N_{\mathrm{charged}}}-<\frac{Q}{N_{\mathrm{charged}}}>\right) ^{2}>
\end{equation}
 where \( Q=N_{+}-N_{-} \) is the charge in the box. The motivation for choosing
these ratios was to reduce the dependence of multiplicity fluctuations caused
by the event structure.

\section{{\large Evaluation of the measures}\large }

In the region of interest {\small for large nuclei at high energies an}d \emph{strong
centrality} the charge component {\small of the fluctuations dominates. In these
region all measures are simply connected b}y the following relations~\cite{jeon00}:
\begin{equation}
\label{eq-7}
<N_{\mathrm{charged}}><\delta R^{2}>=4\: <N_{\mathrm{charged}}><\delta F^{2}>=4\cdot \frac{<\delta Q^{2}>}{<N_{\mathrm{charged}}>}.
\end{equation}

{\par\raggedright and the question of the optimal quantity is somewhat esoteric.
To show this statement all three quantities were calculated in the Dual Parton
model implementation DPMJET~\cite{DPMJET} shown in figure~3. 
\begin{figure*}
{\par\centering \resizebox*{0.48\columnwidth}{0.4\textheight}{\includegraphics{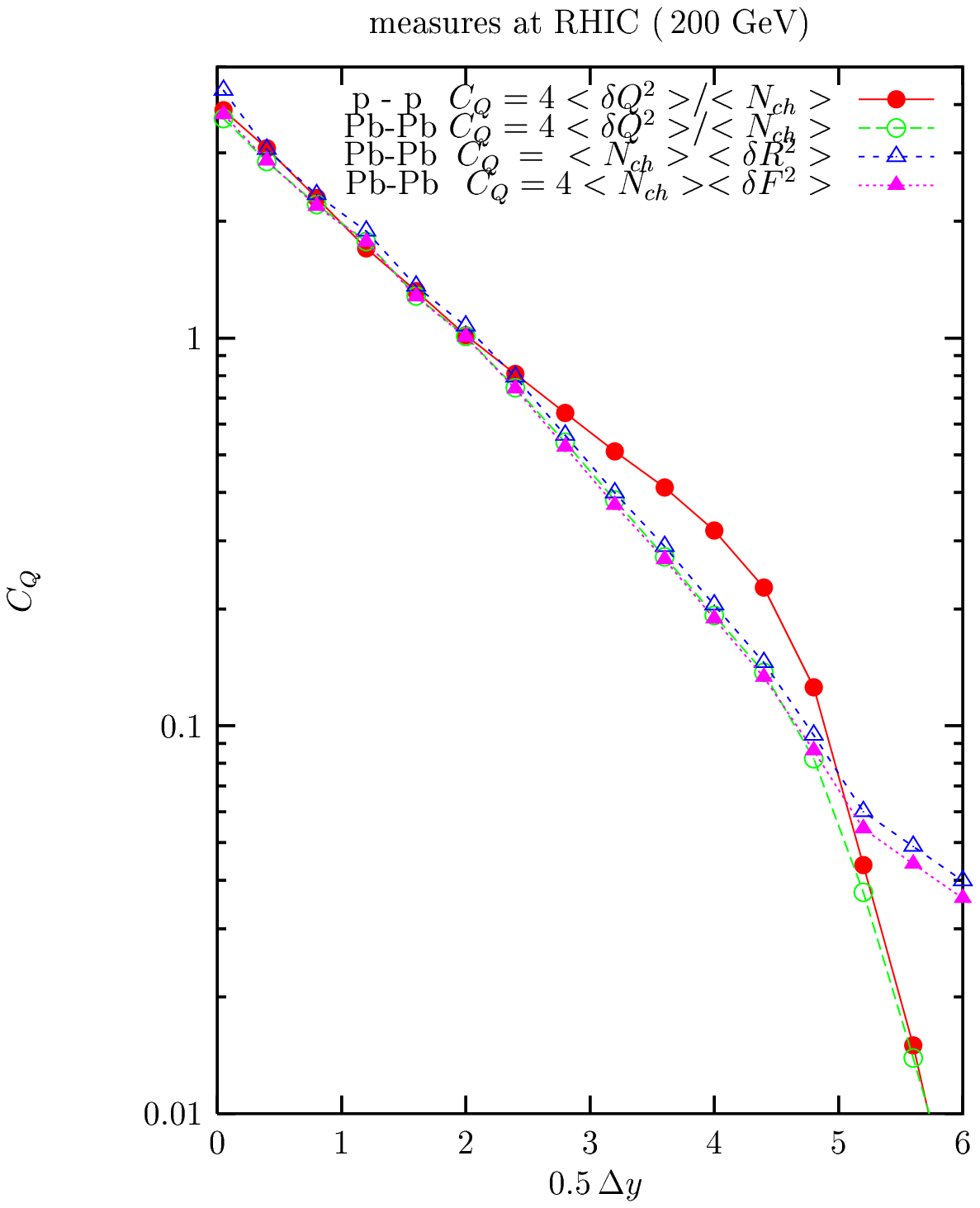}} 
\resizebox*{0.48\columnwidth}{0.4\textheight}{\includegraphics{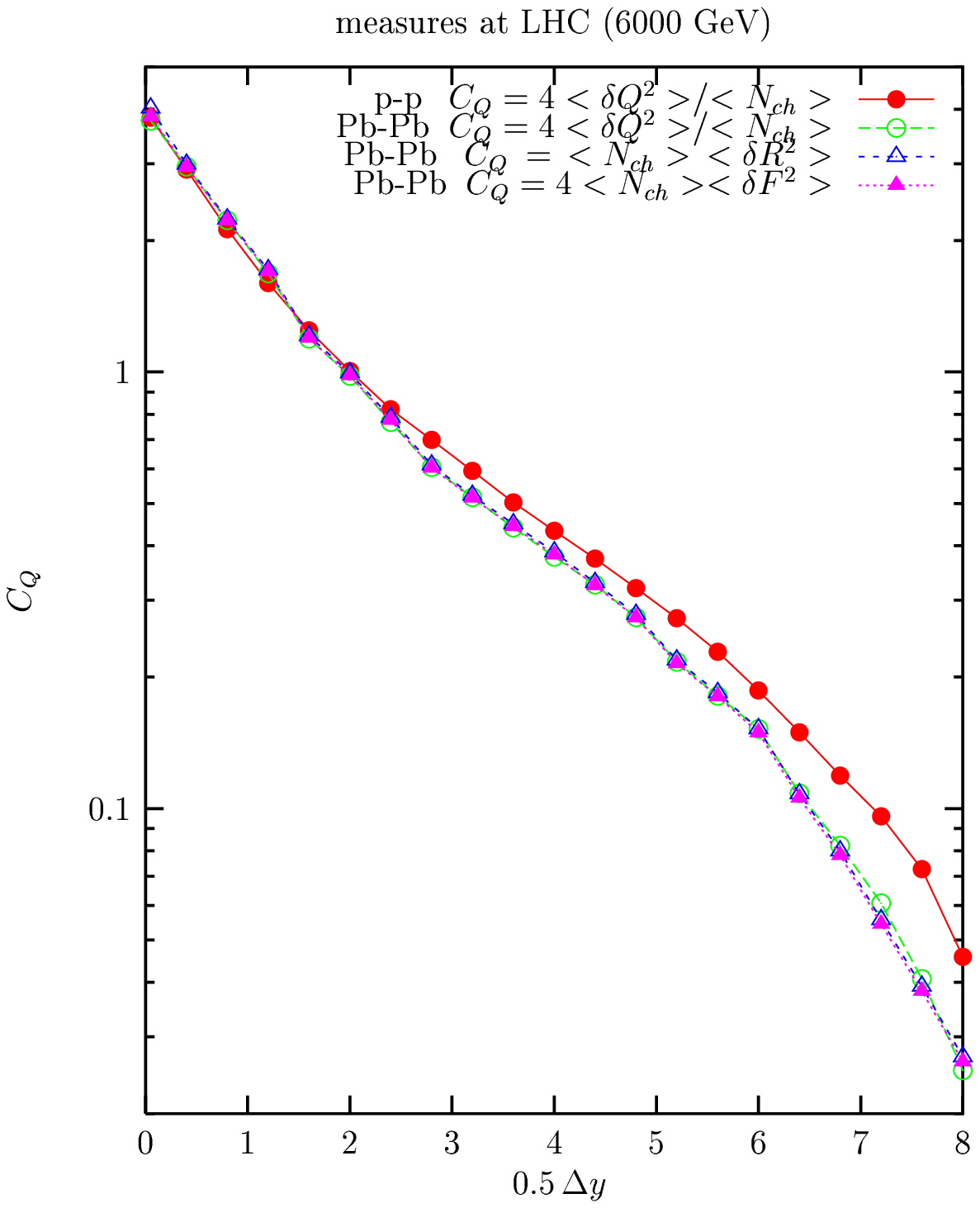}} \par}

\caption{Charge fluctuations for the most central \protect\( 5\%\protect \) Pb-Pb scattering
at RHIC energies (\protect\( \sqrt{s}=200\protect \) A GeV and at LHC energies
(\protect\( \sqrt{s}=6000\protect \) A GeV). Also shown are corresponding data
for p-p scattering}
\end{figure*}
 \par}

{\par\raggedright For the most central \( 5\% \) Pb-Pb scattering at LHC energies
(\( \sqrt{s}=6000 \) A GeV) there is indeed a perfect agreement between all
three quantities as shown in the figure. This agreement stays true for analogous
Pb-Pb data at RHIC energies (\( \sqrt{s}=200 \) A GeV) . \par}

Outside the region of interest - i.e. in the region of \emph{lower particle
densities} - the conventional dispersion, \( <\delta Q> \), has clear advantages.
The alternatives are not suitable for small \( \Delta y \) boxes in less dense
events, 

\begin{itemize}
\item as if no particle in the corresponding box exists in rare events \( 0/0 \)
or \( \infty  \) is undefined and 
\item as if one somehow fixes the problem (e.g. by not considering problematic events)
their mutual relation is destroyed. For the minimum bias S-S scattering at these
energies the agreement is lost and the new measures behave rather erratic~\cite{initial}
. The same erratic behavior for the new measures is found for the proton-proton
case (figure ~4). 
\begin{figure}
{\par\centering \resizebox*{0.6\textwidth}{!}{\includegraphics{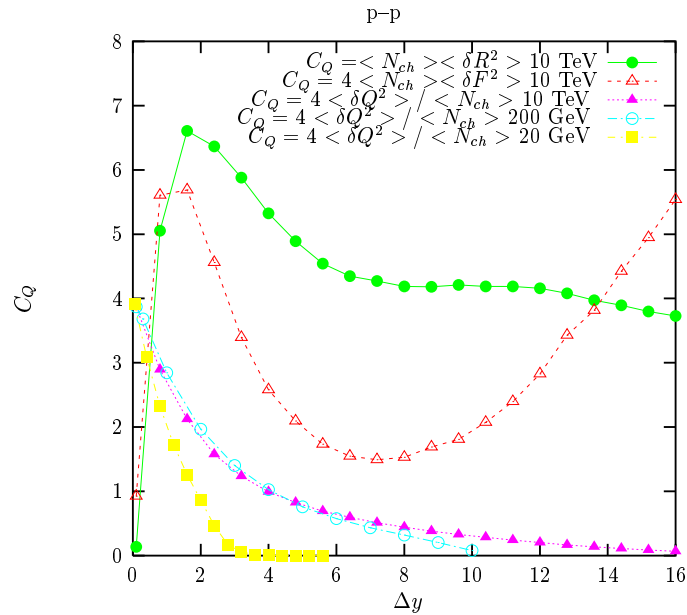}} \par}

\caption{Charge fluctuations for minimum bias pp scattering at SPS, RHIC and LHC energies}
\end{figure}

\end{itemize}
As any conclusion will have to depend on a comparison of central processes with
minimum bias and proton-proton events, there is a clear advantage to stick to
the dispersion of the net charge distribution \( <\delta Q^{2}> \)\footnote{
New RHIC data of the PHENIX collaboration \cite{phenix02} which appeared after
the talk resp. paper\cite{bopp01} confirm the problem with \( <\delta R^{2}> \)
which does not appear for \( <\delta Q^{2}> \). Equation 7 does not hold in
their case as a restricted azimuthal range was considered effectively reducing
the density. For the measured very narrow rapidity range the data are essentially
consistent with (hadronic) statistical fluctuations.
} .

\section{{\large A Simple Relation between the Quark Line Structure and Fluctuations
in the Charge Flow}\large }

To visualize the meaning of charge flow measurements it is helpful to introduce
a general factorization hypothesis. It postulates that the light flavor structure
of an arbitrary hadronic amplitude can be described simply by an overall factor,
in which the contribution from individual quark lines factorize. It is for most
purposes (which consider long range fluctuations) an adequate approximation. 

The hypothesis can be used to obtain the following \emph{generalization of the
Quigg-Thomas relation}~\cite{baier74,aurenche77,bopp78}. It states that the
correlation of the charges \( Q(y_{1}) \) and \( Q(y_{2}) \), which are exchanged
during the scattering process across two kinematic boundaries, is just

{\par\centering {\small 
\begin{equation}
\label{eq-8}
<\delta Q(y_{1})\cdot \delta Q(y_{2})>=n_{\mathrm{common}\: \mathrm{lines}}<\delta q^{2}>.
\end{equation}
}\small \par}

\begin{itemize}
\item {\small where the charges \( \delta Q(y_{i})=Q(y_{i})-<Q(y_{i})> \) were exchanged
across two kinematic boundaries \( y_{1} \)~\&~\( y_{2} \), }{\small \par}
\item {\small where \( q \) is the charge of the quark on such a line. Values \( <\delta q^{2}>=<(q-<q>)^{2}=0.22\cdots 0.25 \)
are obtained, and} 
\item {\small where \( n_{\mathrm{common}\: \mathrm{lines}} \) counts the number
of quark lines intersecting both borders, as illustrated in the simple example
given in figure~5:}
\begin{figure}
{\par\centering \resizebox*{0.3\textwidth}{!}{\includegraphics{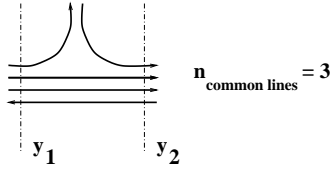}} \par}

\caption{Example of a quark line graph}
\end{figure}
{\small \par}
\end{itemize}
Most observables of charge fluctuations can be expressed using this basic correlation.
Our fluctuation of the charges within a \( [-y_{\mathrm{max}.},+y_{\mathrm{max}.}] \)
box contains a combination of three such correlations. A simple summation yields:
\begin{equation}
\label{eq-9}
<\delta Q[\mathrm{box}]^{2}>=n_{\mathrm{lines}\: \mathrm{entering}\: \mathrm{box}}<(q-<q>)^{2}
\end{equation}
where \( n_{\mathrm{lines}\: \mathrm{entering}\: \mathrm{box}} \) is the number
of quark lines entering the box.

\section{{\large Charge Fluctuations in Equilibrium Models}\large }

Let us use this relation to consider the prediction in more detail. In the \emph{thermalized
limit} with an infinite reservoir outside and a finite number of quarks inside,
all quark inside will connect with quark lines to the outside as shown in figure~6.
The dispersion of the charge transfer is therefore proportional to the total
number of quarks or particles inside. 

In an ``\emph{hadron gas}'' all particles contain two independent quarks each
contributing to the fluctuation with roughly \( 1/4 \) yielding the estimate
\( 1/2 \) as required by equation~2 . In the factorizing limit mesons have
a \( 50\% \) chance to be charged; possible baryon contribution require only
a minor correction. 

For the ``\emph{quark gluon gas}'' ignoring hadronization one obtains one
quark charge fluctuation \( 1/4 \) for each charged parton. Equation~2 is drastically
changed by a factor of \( 4 \). It is, however, not easy for this prediction
to survive hadronization. If hadronization would just group initial partons
into hadrons, the factorizing hadron gas description would stay completely unchanged.
For the reduction it is essential to have only a single quark line contributing
to the fluctuation. Only one quark of each hadron has to originate in the primary
partonic process and the other quark has to originate in local fluctuations
and has not to contribute. The mechanism requires a sufficiently large box so
that short range correlation can be avoided. 
\begin{figure}
{\par\centering \resizebox*{!}{0.09\textheight}{\includegraphics{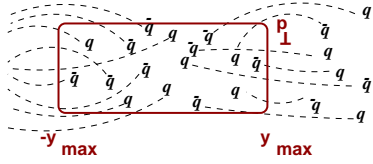}} \par}

\caption{Quark lines entering the box in the thermodynamic limit }
\end{figure}

\section{{\large The expanding Box}\large }

Let us first consider the \emph{limit of a tiny box}. Looking only at the first
order in \( \Delta y \) one trivially obtains in any model

\begin{equation}
\label{eq-10}
<\delta Q^{2}>/<N_{\mathrm{charged}}>=1
\end{equation}
which corresponds to the hadron gas value. If the box size increases to one
or two units of rapidity on each side this ratio will typically decrease, as
most models contain a short range component in the charge fluctuations. The
decreasing is not very distinctive. In hadron hadron scattering processes such
short range correlations are known to play a significant role and there is no
reason not to expect such correlations for the heavy ion case.

After a box size passed the short range the decisive region starts. In all global
equilibrium models~\cite{heinz00,jeon99,jeon00} the ratio will have to reach
now a flat value. The only correction comes from overall charge conservation.
If the box involves a significant part of the total rapidity, it will force
the ratio to drop by a correction factor

\begin{equation}
\label{eq-11}
\mathrm{factor}=\left( \int _{y_{\mathrm{max}.}}^{Y_{\mathrm{kin}.\mathrm{max}.}}\rho _{\mathrm{charge}}^{\mathrm{new}}dy\right) /\left( \int _{0}^{Y_{\mathrm{kin}.\mathrm{max}.}}\rho _{\mathrm{charge}}^{\mathrm{new}}dy\right) \propto 1-y_{\mathrm{max}.}/Y_{\mathrm{kin}.\mathrm{max}.}
\end{equation}
.

\section{{\large Charge Fluctuations in String Models}\large }

This flatness is not expected in string models and numerical calculations indicate
a manifestly different behavior. Only quark lines which intersect boundaries
and which contribute to the charge measure have to be considered. String models
contain local compensation of charge. Only contributions of lines originating
around the boundaries (as illustrated in the figure~7) will appear. 
\begin{figure}
{\par\centering \resizebox*{!}{0.09\textheight}{\includegraphics{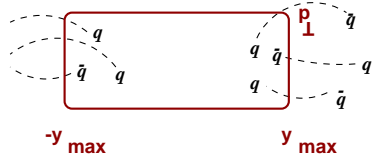}} \par}

\caption{Quark lines entering the box with local compensation of charge}
\end{figure}
 If the distance is larger than the range of charge compensation the dispersion
will no longer increase with the box size. The total contribution will now be
just proportional to the \emph{density of the particles at the boundaries}:
\begin{equation}
\label{eq-12}
<\delta Q^{2}>\, \propto \rho _{\mathrm{charged}}(y_{\mathrm{max}.}).
\end{equation}
 It now just counts the number of strings.

This resulting scaling is illustrated in a comparison between both quantities
in (12). Shown in figure~8 are the predictions of the Dual Parton model implementation
DPMJET~\cite{DPMJET} for RHIC and LHC energies.
\begin{figure}
{\par\centering \resizebox*{!}{0.42\textheight}{\includegraphics{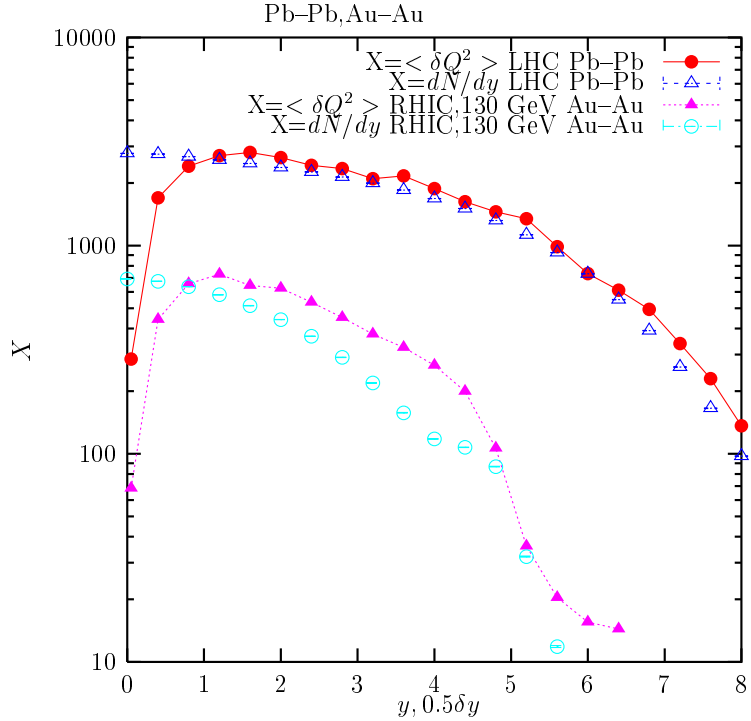}} \par}

\caption{Comparison of the dispersion of the charge distribution with the density on
the boundary of the considered box for central gold gold resp. lead lead scattering
at RHIC and LHC energies.}
\end{figure}
 The agreement is comparable to the proton-proton case shown in figure~9.
\begin{figure}
{\par\centering \resizebox*{!}{0.4\textheight}{\includegraphics{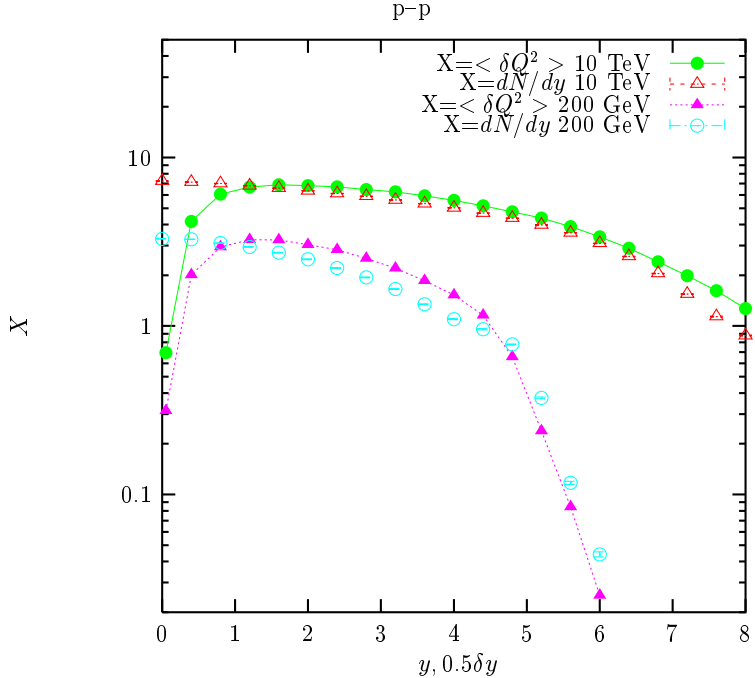}} \par}

\caption{Comparison of the dispersion of the charge distribution with the density on
the boundary of the considered box for proton-proton scattering at RHIC and
LHC energies.}
\end{figure}
 The proportionality is expected to hold for a gap with roughly \( \frac{1}{2}\delta y>1 \)
as for smaller boxes some of the quark lines intersect both boundaries. For
large rapidity sizes there is a minor increase from the leading charge flow
\( Q_{L} \) originating in the incoming particles. In a more careful consideration~\cite{bopp78}
one can subtract this contribution
\begin{equation}
\label{eq-13}
<\delta Q^{2}>_{\mathrm{leading}\: \mathrm{charge}\: \mathrm{migration}}=<Q_{L}>(1-<Q_{L}>)
\end{equation}
 and concentrate truly on the fluctuation. 
\vspace{0.3cm}

A rough estimate of the relative size - with a width of neighboring string break
ups and a width from resonance decays - leads to consistent values~\cite{bopp01}.

\section{{\large Bleicher, Jeon, Koch's observation }\large }

In a recent publication Bleicher, Jeon, Koch~\cite{bleicher00} showed:

\begin{itemize}
\item The overall charge conservation cannot be ignored at SPS energies 
\item It obliviates in this energy range the distinction between even the most extreme
models including string models and statistical models with hadronic equilibrium. 
\end{itemize}
They showed that their string model prediction\footnote{
In the energy above \( \sqrt{s}=5 \) GeV their UrQMD code is described~\cite{bleicher99}
to be dominated by string fragmentation. 
} coincides with the expectation of a statistical model of hadrons. Our string
model DPMJET supports this conclusion for the SPS energy range as it also obtains
fluctuations consistend with the ``statistical'' expectation. 

While forward-backward hemisphere charge fluctuations were meaningful in the
FNAL-SPS energy region, the fluctuations of charges into a central box contain
two borders and require a correspondingly doubled rapidity range. Unfortunately
this means a lot in energies. They are not available at SPS energies.

\section{{\large A reference model with statistical fluctuation}\large }

It was argued~\cite{bleicher00} that the experimental results should be \char`\"{}purified\char`\"{}
to account for charge conservation. We basically aggree with such a correction.
Given the uncertainites the correction obviously has to stay on the modelling
side.

{\par\raggedright To obtain such a reference model we \emph{randomize charges}
\emph{a posteriori}. To accurately conserve energy and momentum it was done
separately for pions, kaons and nucleons\footnote{
Obviously, the method can also be directly applied to experimental data, at
least in a simplified way. 
}. Using DTMJET events for RHIC and \textbf{}LHC energies for \emph{proton-proton}
and \emph{central lead-lead} collisions we obtain the ``statistical'' prediction
shown in figure~10. \par}

To check consistency we employed the correction factor proposed by ~\cite{bleicher00}
\[
1-\int _{0}^{y_{\mathrm{max}.}}\rho _{\mathrm{charge}}^{}\: dy/\int _{0}^{Y_{\mathrm{kin}.\mathrm{max}.}}\rho _{\mathrm{charge}}^{}\: dy\]
and indeed obtained the flat distribution with the expected ``hadron gas''
value. 

\begin{figure}
{\par\centering \resizebox*{0.48\columnwidth}{0.42\textheight}{\includegraphics{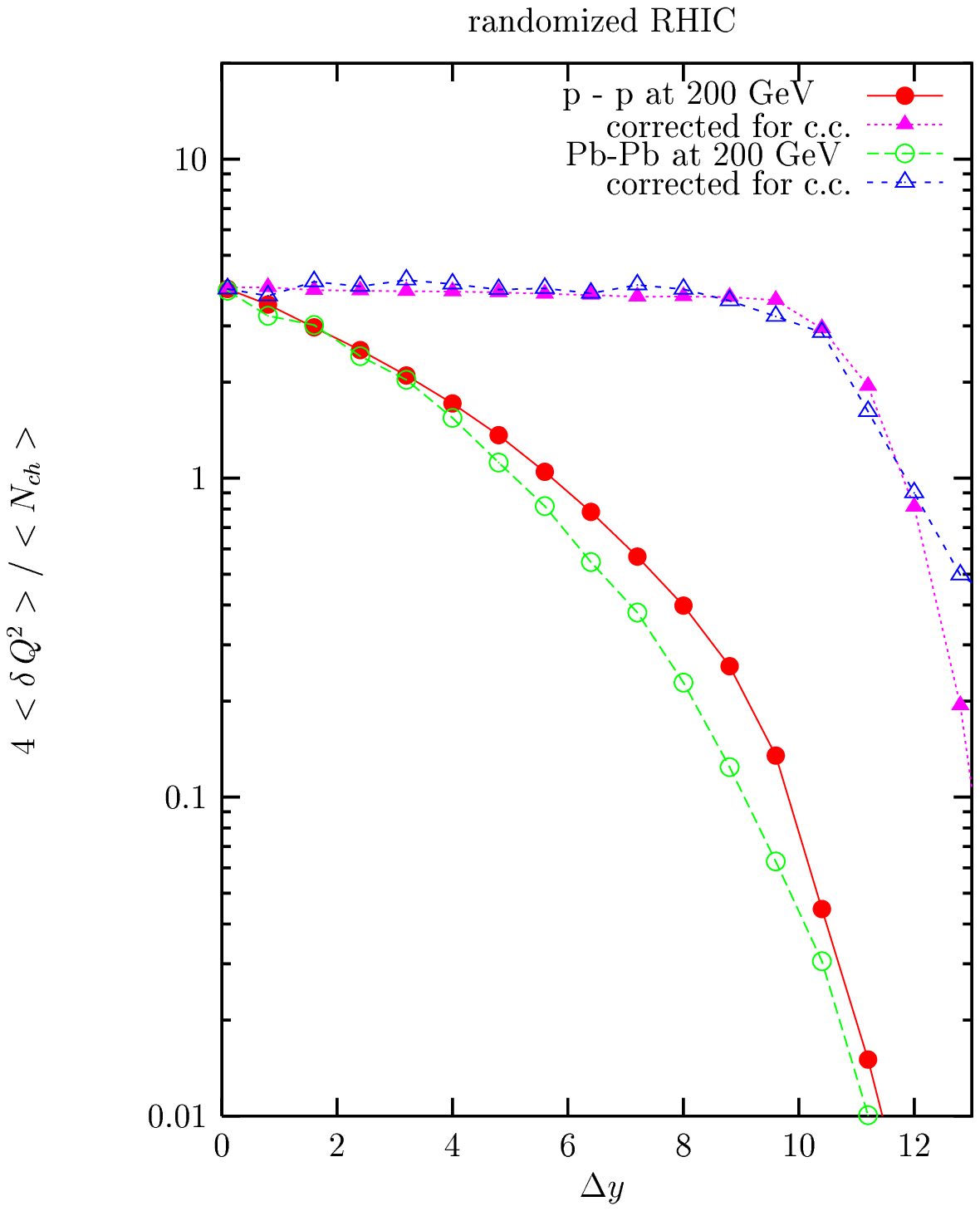}} 
\resizebox*{0.48\columnwidth}{0.42\textheight}{\includegraphics{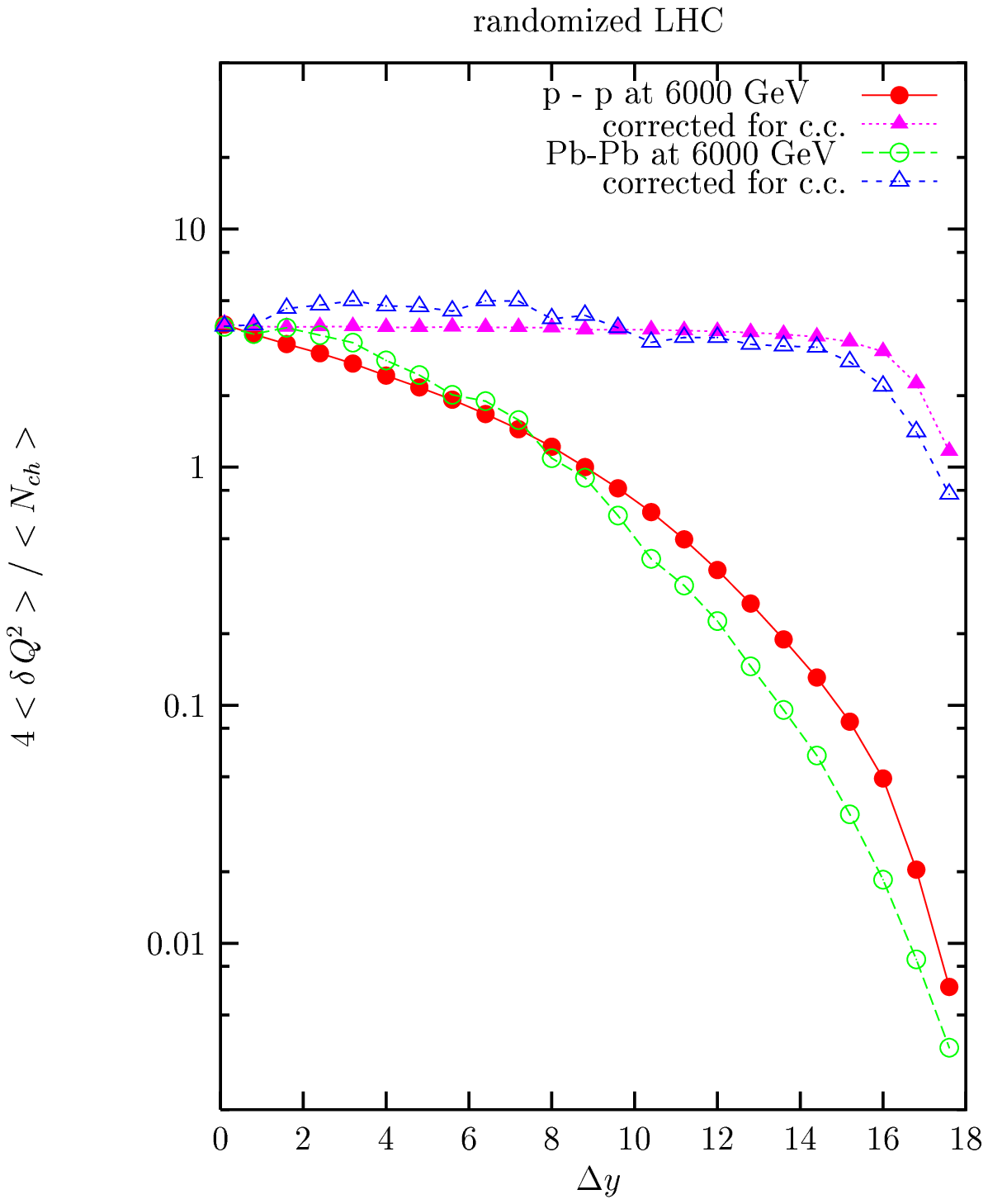}} \par}

\caption{Charge fluctuations with a posteriori randomized charges for p-p scattering
and the most central \protect\( 5\%\protect \) in Pb-Pb scattering at RHIC
energies (\protect\( \sqrt{s}=200\protect \) A GeV) and at LHC energies (\protect\( \sqrt{s}=6000\protect \)
A GeV). The results are also shown with a correction factor to account for the
overall charge conservation.}
\end{figure}

\section{{\large String model versus randomized ``hadron gas'' }\large }

Taking the DPMJET string model and the randomized ``hadron gas'' version as
extreme cases (with the ``parton gas'' somewhere in between) we can investigate
the decisive power of the measure. As shown in figure~11 we find that there
is a measurable distinction at RHIC energies and sizable at LHC energies. 
\begin{figure}
{\par\centering \resizebox*{0.48\columnwidth}{0.35\textheight}{\includegraphics{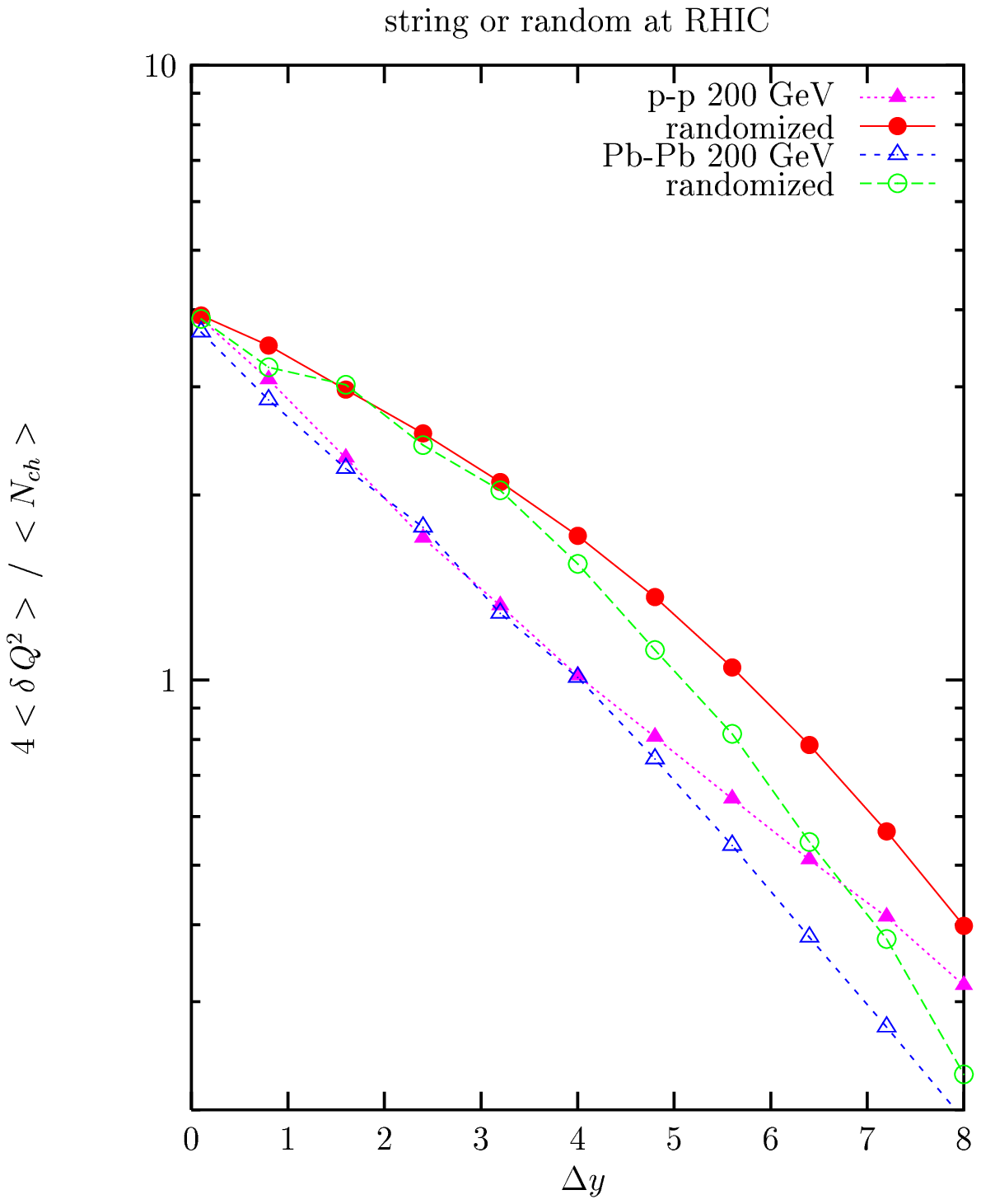}} 
\resizebox*{0.48\columnwidth}{0.35\textheight}{\includegraphics{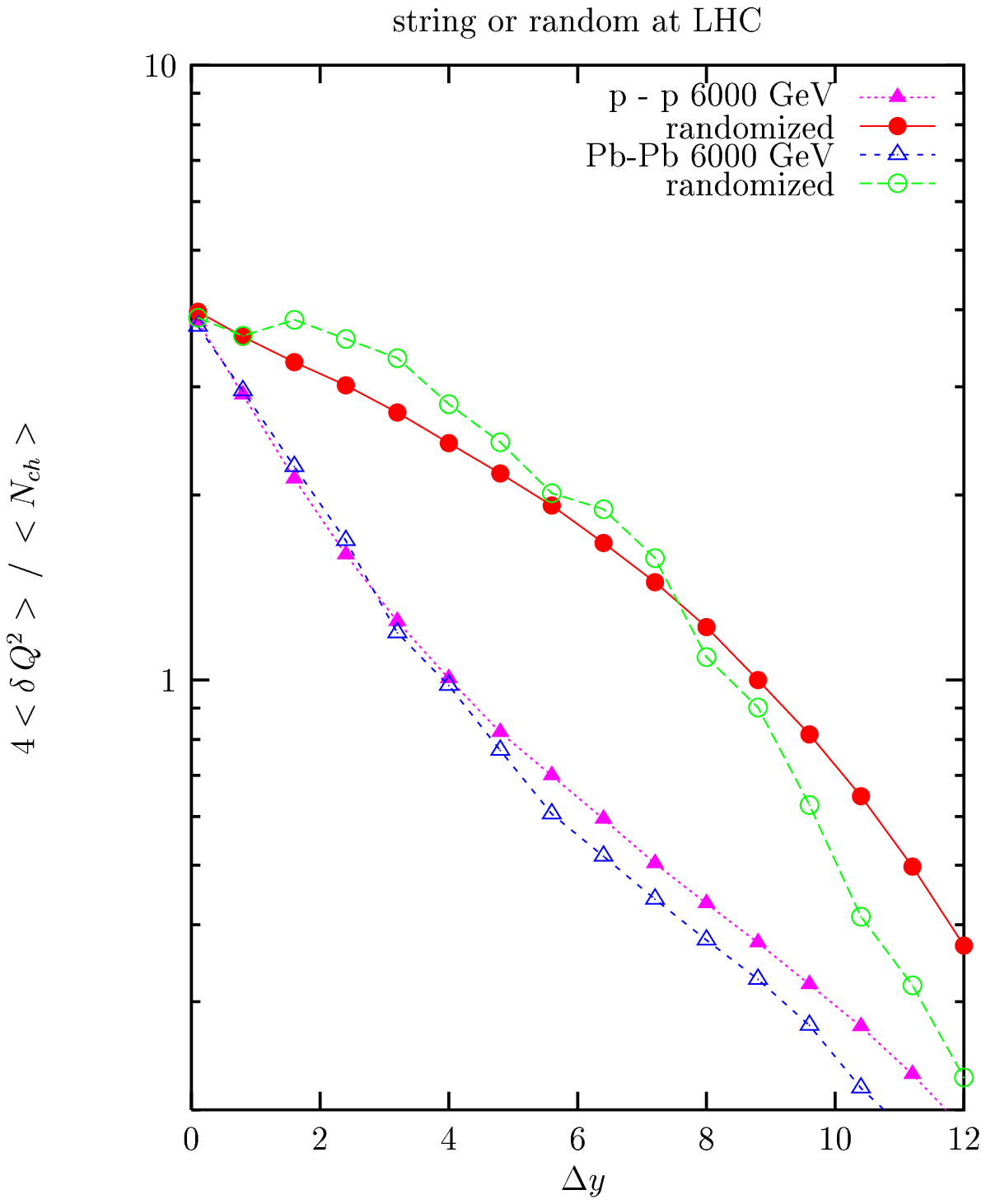}} \par}

\caption{Comparison of the charge fluctuations obtained in a string model DPMJET with
a model using a posteriori randomized charges for p-p scattering and the most
central \protect\( 5\%\protect \) in Pb-Pb scattering at RHIC energies (\protect\( \sqrt{s}=200\protect \)
A GeV) and at LHC energies (\protect\( \sqrt{s}=6000\protect \) A GeV). }
\end{figure}

The similarity of p-p and Pb-Pb scattering is not surprising. The distinction
between both cases is expected from the difference in collective effects. The
data for p-p scattering are known to follow the string models, while interaction
of comovers, or medium range or complete equilibrium will move the curve upward
to a more statistical situation. These effects are presently outside of the
model. A measured charge correlation between both extremes will directly reflect
the underlying new physics.

\section{{\large The b dependence of the charge fluctuations}\large }

A similar result is obtained when the dependence on the centrality is studied. 
\begin{figure}
{\par\centering \resizebox*{0.6\columnwidth}{0.4\textheight}{\includegraphics{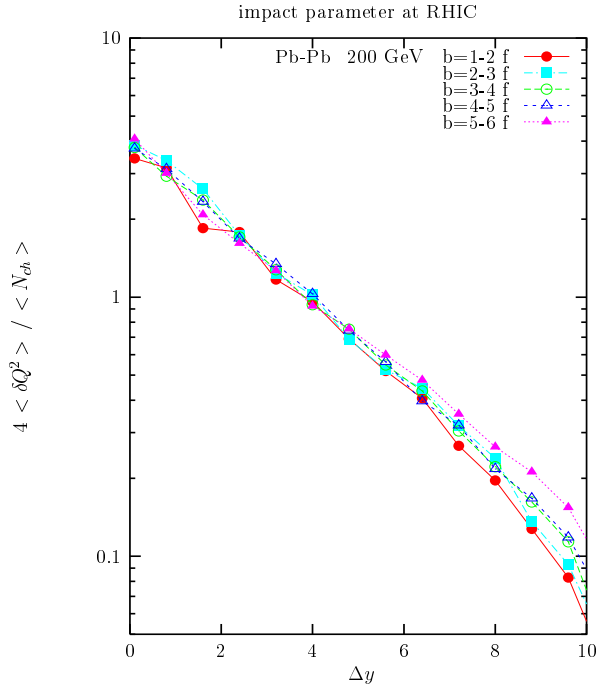}} \par}

\caption{The b dependence of the charge fluctuations obtained in a string model DPMJET
for p-p scattering and the most central \protect\( 5\%\protect \) in Pb-Pb
scattering at RHIC energies (\protect\( \sqrt{s}=200\protect \) A GeV). }
\end{figure}
 Without collective effects no such dependence is expected as observed in the
model calculation shown in figure~12 (\( b \) is the impact parameter). This
experimentally measurable centrality dependence allows to directly observe collective
effects without reference to model calculations and underlying concepts.

\section*{{\large Conclusion}\large }

In the paper we demonstrated that the dispersion of the charge distribution
in a central box of varying size is an extremely powerful measure. 

Within the string model calculation the dispersion seen in relation to the spectra
shows no difference between simple proton-proton scattering and central lead
lead scattering even though both quantities change roughly by a factor of 400. 

The dispersion allows to clearly distinguish between conventional string models
and hadronic thermal models for a rapidity range which could be available at
RHIC energies. In many models the truth is expected to lie somewhere in between
and it is a reasonable expextation that the situation can be positioned in a
quantitative way.


\begin{thebibliography}{10}
\bibitem{idschok73}U.~Idschok et al. [Bonn-Hamburg-Munich Collaboration], Nucl.\ Phys.\ {\bf B67}, 93 (1973). 
\bibitem{bialas75}A.~Bialas, K.~Fialkowski, M.~Jezabek and M.~Zielinski, Acta Phys.\ Polon.\ B {\bf 6}, 39 (1975). 
\bibitem{whitmore76}J.~Whitmore, Phys.\ Rept.\ {\bf 27}, 187 (1976). 
\bibitem{foa75}L.~Foa, Phys.\ Rept.\ {\bf 22}, 1 (1975). 
\bibitem{brandelik81}R.~Brandelik {\it et al.} [TASSO Collaboration], Phys.\ Lett.\ B {\bf 100}, 357 (1981).
\bibitem{quigg73}C.~Quigg and G.~H.~Thomas, Phys.\ Rev.\ {\bf D7},2757 (1973). 
\bibitem{quigg75}C.~Quigg, ` Phys.\ Rev.\ {\bf D12}, 834 (1975).
\bibitem{bopp78}F.~W.~Bopp, Riv.\ Nuovo Cim.\ {\bf 1}, 1 (1978). 
\bibitem{misra83}S.~P.~Misra and B.~K.~Parida, Pramana{\bf 20}, 375 (1983).
\bibitem{ranft75}J.~Ranft, Fortsch.\ Phys.\ {\bf 23}, 467 (1975).
\bibitem{chouyang73}T.~T.~Chou and C.~N.~Yang, ~Phys.\ Rev.\ {\bf D7}, 1425 (1973).
\bibitem{sivers74}J.~L.~Newmeyer and D.~Sivers, Phys.\ Rev.\ D {\bf 10}, 204 (1974). 
\bibitem{phua77}K.~F.~Loe, K.~K.~Phua and S.~C.~Chan, Lett.\ Nuovo Cim.\ {\bf 18}, 137 (1977). 
\bibitem{lam77}E.~N.~Argyres and C.~S.~Lam, Phys.\ Rev.\ D {\bf 16}, 114 (1977). 
\bibitem{chiu76}C.~B.~Chiu and K.~Wang, Phys.\ Rev.\ D {\bf 13}, 3045 (1976). 
\bibitem{diasdedeus77}J.~Dias de Deus and S.~Jadach, Phys.\ Lett.\ B {\bf 66}, 81 (1977). 
\bibitem{jezabek77}M.~Jezabek, Phys.\ Lett.\ B {\bf 67}, 292 (1977).
\bibitem{heinz00}M.~Asakawa, U.~Heinz and B.~Muller, Phys.\ Rev.\ Lett.\ {\bf 85}, 2072 (2000).
\bibitem{jeon00}S.~Jeon and V.~Koch, Phys.\ Rev.\ Lett.\ {\bf 85}, 2076 (2000) [hep-ph/0003168].
\bibitem{jeon99}S.~Jeon and V.~Koch, Phys.\ Rev.\ Lett.\ {\bf 83}, 5435 (1999)[nucl-th/9906074].
\bibitem{bleicher00}M.~Bleicher, S.~Jeon and V.~Koch, Phys.\ Rev.\ {\bf C62}, 061902 (2000) [hep-ph/0006201];
V.~Koch, M.~Bleicher and S.~Jeon, nucl-th/0103084.
\bibitem{bopp81}F.~W.~Bopp,Nucl.\ Phys.\ {\bf B191}, 75 (1981). 
\bibitem{bialas02}A.~Bialas, arXiv:hep-ph/0203047.
\bibitem{fialkowski00}K.~Fialkowski and R.~Wit,hep-ph/0006023, ``Charge fluctuations in a final state with QGP,''hep-ph/0101258.
\bibitem{DPMJET}J.~ Ranft, Phys.\ Rev. {\bf D 510} 64 (1995); J.~Ranft, hep-ph/9911213 (Siegen preprint SI-99-5); J.~Ranft, hep-ph/9911232 (Siegen preprint SI-99-6).
\bibitem{initial}Figure 2 of initial version: F.~W.~Bopp and J.~Ranft, hep-ph/0009327. 
\bibitem{phenix02}K.~Adcox [PHENIX Collaboration], arXiv:nucl-ex/0203014. 
\bibitem{bopp01}F.~W.~Bopp and J.~Ranft, Eur.\ Phys.\ J.\ C {\bf 22}, 171 (2001) arXiv:hep-ph/0105192.
\bibitem{baier74}R.~Baier and F.~W.~Bopp, Nucl.\ Phys.\ {\bf B79}, 344(1974).
\bibitem{aurenche77}P.~Aurenche and F.~W.~Bopp, Nucl.\ Phys.\ {\bf B119}, 157 (1977).
\bibitem{bleicher99}M.~Bleicher {\it et al.}, J.\ Phys.\ {\bf G25} (1999) 1859 [hep-ph/9909407]. 
\end{thebibliography}
\end{document}